\documentclass[preprint,12pt]{elsarticle}

\usepackage{amssymb, amsmath}
\usepackage{xcolor}

\usepackage{optidef}

\usepackage{tikz, pgfplots}
\pgfplotsset{compat=newest}

\usepackage{graphicx}
\usepackage{subcaption}

\usepackage{algorithm}
\usepackage{algpseudocode}

\usepackage{url}

\usepackage{gensymb}

\usepackage{makecell}

\journal{Smart Energy}

\begin{document}

\begin{frontmatter}



\title{A data-based comparison of methods for reducing the peak volume flow rate in a district heating system}


\author[inst1]{Felix Agner}

\affiliation[inst1]{organization={Department of Automatic Control, Lund University},
            addressline={Ole Römers väg 1}, 
            city={Lund},
            postcode={223 63}, 
            country={Sweden}}

\author[inst2]{Ulrich Trabert}
\author[inst1]{Anders Rantzer}
\author[inst2]{Janybek Orozaliev}

\affiliation[inst2]{organization={Institute of Thermal Engineering, University of Kassel},
            addressline={Kurt-Wolter-Str. 3}, 
            city={Kassel},
            postcode={34125}, 
            country={Germany}}

\begin{abstract}
This work concerns reduction of the peak flow rate of a district heating grid, a key system property which is bounded by pipe dimensions and pumping capacity. The peak flow rate constrains the number of additional consumers that can be connected, and may be a limiting factor in reducing supply temperatures when transitioning to the 4th generation of district heating. We evaluate a full year of
operational data from a subset of customer meters in a district heating system in Germany. We consider the peak flow rate reduction that could be achieved with full a posteriori knowledge of this data. Three strategies for reducing the peak flow rate are investigated: 
A load shifting demand response strategy, 
 an upper limitation in substation return temperatures, and an upper limitation on each substation's volume flow rate.
We show that imposing up to to 18 \% load flexibility for the customers provides an equal reduction in the peak system flow rate under the load shifting strategy. The limited return temperature strategy is less efficient at curtailing the peak flow rate, but provides an overall reduction of volume flow rates. Finally, the flow rate limitation method can introduce new, higher flow rate peaks, reducing performance.

\end{abstract}

\begin{keyword}
District Heating \sep Demand Response \sep Optimization \sep Smart Meter Data \sep Peak Flow Rate Reduction
\end{keyword}

\end{frontmatter}


\section{Introduction}
\label{sec:introduction}

In the smart energy systems of the future, efficient system designs involve heavy couplings between energy sectors \cite{Lund.2017}. A key component of this future energy system is the 4th generation of district heating systems \cite{lund_4th_2014}. Part of the reason why 4th generation district heating is a key component is the flexibility that these systems can provide \cite{Vandermeulen.2018}, along with the ability to integrate renewable heat sources and low-temperature heat sources such as waste heat \cite{lund_4th_2014}. One of the key differences between 4th generation and former generations of district heating is that the supply temperature in the 4th generation of systems is much lower, enabling the use of renewable and low-temperature heat sources.

The transition to lower temperature systems brings several challenges with it \cite{GUELPA.2023.supply-temp-review}. In particular, challenges can arise when retrofitting existing district heating grids for lower temperatures \cite{GUELPA.2023.supply-temp-review, Brange.2017}. This is in part due to the fact that lowering the supply temperature in the grid may not lead to an equivalent reduction in return temperature. This corresponds to a reduction in $\Delta T$ at the customer substation. Hence, in order to supply the same heat load, a corresponding increase in volume flow rates will occur. Higher system flow rates can incur several practical issues, among which are higher pumping costs \cite{GUELPA.2023.supply-temp-review} and bottleneck scenarios \cite{Brange.2017}, which corresponds to cases where peripheral customers experience low differential pressures due to high pressure losses in the pipes. Several ways to combat the negative impacts of lowering supply temperature have been considered in the literature \cite{GUELPA.2023.supply-temp-review, Brange.2017}, and in this paper we focus on three such strategies in particular. Two of these can be considered instances of \textit{demand response} and one is an instance of \textit{improved cooling in substations}.

Both demand response and improved cooling in substations fall under the umbrella term of \textit{demand side management} \cite{GUELPA2021119440-dsm-review}, meaning interventions aimed at changing the load pattern of consumers. Demand response corresponds to methods where load patterns are directly controlled through e.g. load shedding or load shifting. Better cooling in substations falls under the retrofitting category, requiring improved equipment being installed in consumer substations.

As it is a key factor in reducing system temperatures in existing district heating systems, many studies have focused on approaches to improved cooling in substations. \citet{Zinko.2005} defined the excess flow method, which is a rather simple approach that ranks district heating substations according to their impact on the return temperature in the whole grid. This method has been applied by \citet{Bergstraesser.2021} to three existing district heating systems in Germany in order to identify faults in substations and propose measures for their repair. They found that the costs of fixing the faults can vary widely, but that it is economically viable for grids with temperature-sensitive heat generation technologies. The costs stated in this study can be compared to the monetary benefits of lower system temperatures, which are presented as cost reduction gradients by \citet{Geyer.2021} for several heat generation technologies. They concluded that the cost reduction gradients for traditional heat generation in combined heat and power plants or heat-only boilers are much lower than for emerging renewable technologies such as heat pumps, geothermal heat, waste heat or solar thermal collectors, highlighting the importance of low system temperatures in future district heating systems.

The reasons for high return temperatures are usually to be found on the secondary side of district heating substations. According to \citet{Tahiri.2023}, one problem is the traditional charging control of domestic hot water tanks, which often leads to overcharging and consequently high return temperatures. Therefore, they propose a novel staged gain control concept, which only requires an additional temperature sensor at the top of the tank and allows a better adjustment of the charging mass flows according to demand. In addition, the buildings' existing radiator heating systems can be optimised to operate at lower temperatures. Based on data from heat cost allocators, \citet{Benakopoulos.2021,Benakopoulos.2022} developed a strategy to minimize the supply temperature in the buildings while maintaining thermal comfort in all apartments. As the design conditions (very cold ambient temperatures) for these systems occur rarely, \citet{Tunzi.2023} state that the radiators are oversized for operation during most of the year, making them suitable for low temperature operation without extensive energy refurbishment of the building.
Besides the local optimization of substation operation, there are also system-level approaches to reducing return temperatures. \citet{Tol.2021} and \citet{VanOevelen.2023} use live data from substations to provide feedback to a central controller that minimizes the required supply temperature in the district heating grid to reduce return temperatures.
Another strategy analysed by \citet{Volkova.2022} is to cascade the heat use by integrating subgrids with lower temperatures into the return line of existing high temperature grids. According to their results, the temperature and flow rates at the connecting point of the subgrid are the crucial parameters to investigate the feasibility of such a concept.

The other considered category of demand side management, demand response, has a rich literature in the domain of power systems and smart grids. Only recently has it gained traction within the domain of district heating. Guelpa et al. \cite{GUELPA2021119440-dsm-review} wrote a review on demand response in district heating in 2021. Already at that time there were several works published on the topic. Guelpa et al. \cite{Guelpa.2019} implemented demand side management in a real district heating system in Turin in 2019. They performed a form of load shifting where they consider shifting load peaks to earlier times for a select number of study participants. They optimized these actions with the goal of reducing system load peaks and showed that they could reduce system peak loads by 5 \% with restrictive shifting policies in the real demonstration site and up to 35 \% with more liberal shifting policies in a simulation environment. Kotila et al. \cite{Kotila.2020} performed field tests in student apartment buildings in Finland, employing a method where space heating is reduced in response to high peaks in domestic hot water usage. After the review of Guelpa et al., several other works on the topic have been published. Van Oevelen et al. \cite{VanOevelen.2023} implemented a demand response approach for reducing the peak energy demand through supply temperature control in a demonstration site in Brescia, Italy under the TEMPO project \cite{tempo_dhc_2023}. Furthermore, Capone et al. performed two simulation studies in 2021 on holistic district heating system optimization, including a load-shifting demand response strategy. They first used this tool to analyze the trade-off between carbon dioxide emissions and economic feasibility \cite{Capone.2021.multi-objective-optimization}, and later analyzed the balance between thermal storage and demand response to show that demand response can provide drastic cost reduction compared to the use of only thermal storage \cite{Capone.2021.storage-and-thermal-dr}.\\

In this paper we employ the following approach to compare the three considered strategies. We will investigate the effectiveness of these strategies when applied retroactively to data from 18 consumer substations from 2022. We aim to answer the following questions: For the district heating system and related data considered in this study, \textit{which considered intervention has the greatest impact on the yearly peak flow rate}, \textit{what is the potential impact of targeting only a subset of consumers with each intervention} and \textit{what level of flexibility is necessary for demand response to be impactful}?

The remainder of this paper is organized as follows. We introduce the three considered strategies in detail, along with a description of the considered data set and preprocessing in section \ref{sec:method}. In section \ref{sec:results} we present the results of the strategies and their impact on the data. We conclude with discussion and future outlooks in section \ref{sec:conclusion}.

\subsection{Notation}
We utilize the notation presented in Table \ref{tab:notation}. In addition, we consider $n=18$ substations, and use subscript $i$ where $i=1,\dots,n$ to denote a value related to substation $i$. We employ the notation $(t)$ to denote a value measured at time instance $t$. For example, $\Delta T_i(t)$ denotes the difference between supply and return temperature at the substation of consumer $i$ measured at hour $t$. We additionally refer to the \textit{aggregate} of a value as the total combined value of all consumers. In particular, the aggregate volume flow rate is simply the sum of the volume flow rates for all consumers. The aggregate return temperature is the flow-rate-weighted sum of all consumers.
\begin{table}[H]
    \centering
    \caption{Notation used in the paper.}
    \begin{tabular}{c|c|c}
        Symbol & Description & Unit\\ \hline
        i & Meter ID & - \\
        $Q$ & Heat energy & kWh \\
        $\Dot{Q}$ & Heat power & kWh/h \\
        RL & Return line & - \\
        SL & Supply line & - \\
        t & Time & h \\
        $T$ & Temperature & $^\circ$C \\
        $\Delta T$ & Temperature difference (supply and return) & $^\circ$C \\
        $\Dot{V}$ & Volume flow rate & m$^3$/h \\
        $\alpha$ & Flexibility level & - \\
        $\beta$ & Flow rate limitation level & - \\
        $\delta$ & Deviation from original volume flow rate & - \\
        $c_p$ & Specific heat capacity of water & kWh/kg$^\circ$C\\
        $\rho$ & Density of water & kg/m$^3$
    \end{tabular}
    \label{tab:notation}
\end{table}

\section{Method}
\label{sec:method}
We investigate a dataset from the full year of 2022 with primary-side substation data from a small set ($n=18$) of consumers in a German district heating system. This entails four time series for each substation sampled at hourly intervals, corresponding to primary side volume flow rates, supply temperatures, return temperatures and heat loads. A more detailed description of the dataset is provided below. For this data, we consider what the impact would have been if three different strategies were employed to reduce the peak flow rate of the system:

\begin{enumerate}
    \item Coordinated optimal load shifting,
    \item individual return temperature limitations and
    \item individual flow rate limitations.
\end{enumerate}
Strategies 1 and 3 correspond to two control strategies where the primary side volume flow rate in each substation is actively regulated, under the assumption that these small changes would not impact the return temperature. Strategy 2 corresponds to an upgrade of the substations, where we instead assume that the return temperature will be affected by the intervention, which subsequently alters the volume flow rates. Each strategy is described in further detail below. Both the coordinated optimal load shifting and the individual flow rate limitations can be considered instances of demand response. Individual return temperature limitations, on the other hand, may need to be supported by additional equipment upgrades to enable better cooling in substations. For each of the three strategies, we define rules for how they impact the behavior of the consumers, and then retroactively apply these rules to the dataset. The altered data set generated by this process then describes the functionality of the system with the strategy in place, which we then compare to the unaltered data. An important note here is that we do not propose specific methods of implementation for the different strategies. The coordinated optimal load shifting strategy would for instance require many technical components in practice. It could be approached with e.g. a centralized model predictive controller and load forecasting, which requires advanced measurement-and-communication devices connected in each substation. Rather than specifying the implementation method in this way, we define the outcome of a particular strategy and quantify the expected impact on the system.

\subsection{Dataset Description and Preprocessing}

The data considered in this research is from 18 smart meters installed on the primary side of district heating substations in an urban area in Nuremberg, Germany. The data points are mean values logged in 5-minute intervals, resampled to hourly mean values, and include volume flow rates, supply and return temperatures, and heat loads. On average, the timeseries in the dataset are more than 93~\% complete. However, in order to avoid distortion of the analysis due to missing datapoints, the gaps in the timeseries are filled using the machine learning methods described by \citet{Trabert.2022}.

\begin{figure}[H]
    \centering
    \includegraphics[width=.5\textwidth]{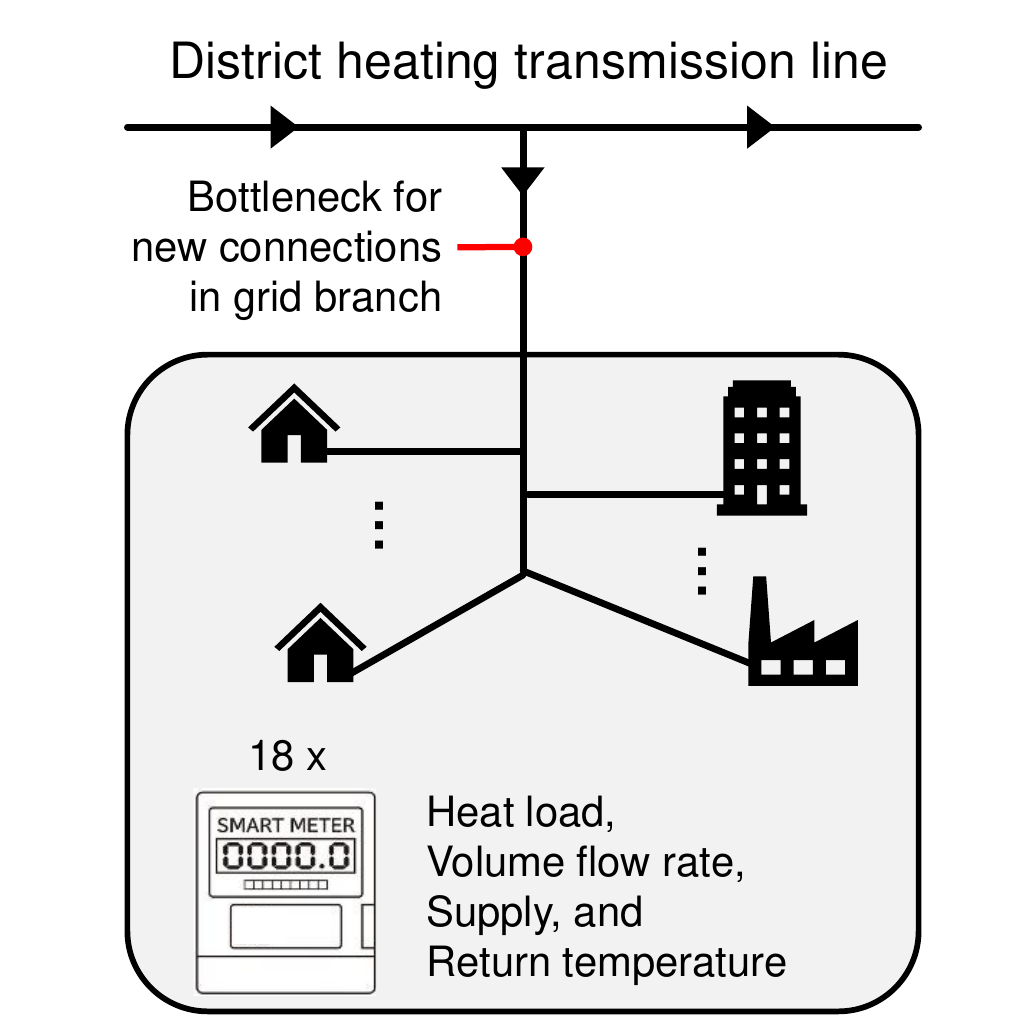}
    \caption{A conceptual scheme of the grid branch where the data used in this study was collected.}
    \label{fig:grid scheme}
\end{figure}

Although the smart meters were only recently installed in 2021, the substations have already been in use for many years. They are located in a branch that is connected to a larger transmission line of the grid and contain residential, commercial, and industrial consumers (see Figure \ref{fig:grid scheme}).

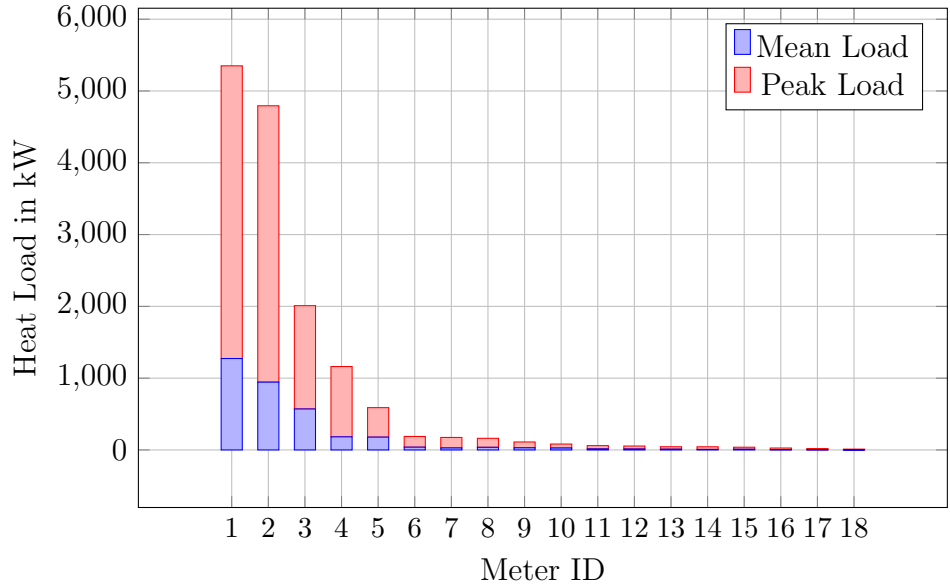
\begin{figure}[H]
    \centering
    \begin{tikzpicture}
\begin{axis}[
    width=.9\textwidth,
    height=.6\textwidth,
    ybar stacked,               
    enlargelimits=0.15,
    ylabel={Heat Load in kW},
    xlabel={Meter ID},
    xtick=data,
    bar width=8pt,
    grid=both,
    legend pos=north east,
    xticklabel style={font=\small},
]
\addplot table[col sep=comma,x=ind,y=mean_load] {data/original_loads_and_deltas.csv};
\addplot table[col sep=comma,x=ind,y=peak_mean_diff_load] {data/original_loads_and_deltas.csv};
\legend{Mean Load, Peak Load}
\end{axis}
\end{tikzpicture}
    \caption{Bar plot of mean and peak heat loads by meter ID.}
    \label{fig:original_load_bars}
\end{figure}

The bar plot of the mean and peak loads of the 18 substations in Figure~\ref{fig:original_load_bars} gives an impression of the consumer structure in the considered branch of the grid. It is dominated by two very large consumers with around 5~MW peak load, which account for about two thirds of the heat consumption in the branch. They are followed by three medium-sized consumers with more than 0.5~MW peak load, while the remaining thirteen substations have only a minor share of around 7~\% in both peak load and heat consumption respectively.

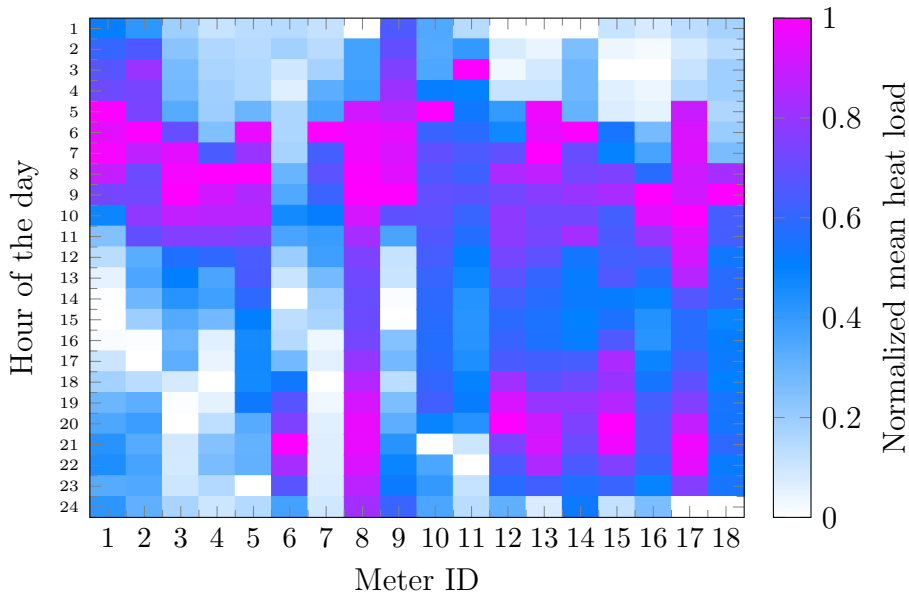
\begin{figure}[H]
    \centering
    \begin{tikzpicture}
    \begin{axis}[
        width=.75\textwidth,
        height=.6\textwidth,
        xlabel={Meter ID},
        ylabel={Hour of the day},
        enlargelimits=false,
        minor tick num=1,
        colorbar,
        colorbar style={
            ylabel={Normalized mean heat load},  
        },
        colormap/cool,
        point meta min=0,
        point meta max=1,
        ymin=0.5,
        ymax=24.5,
        xmin=0.5,
        xmax=18.5,
        mesh/cols=24,
        mesh/rows=18,
        xtick={1,2,3,4,5,6,7,8,9,10,11,12,13,14,15,16,17,18},
        xticklabel style={rotate=0}, 
        ytick={1,2,3,4,5,6,7,8,9,10,11,12,13,14,15,16,17,18,19,20,21,22,23,24},
        xticklabel style={font=\small},
        yticklabel style={font=\tiny},
    ]
        \addplot [matrix plot, point meta=explicit] table [col sep=comma, meta=z, x=x, y=y] {data/original_daily_profiles_matrix.csv};
    \end{axis}
\end{tikzpicture}
    \caption{Mean hourly heat load for each consumer and hour of the day (1-24) taken over the full year of data. The profile for each consumer is normalized in the interval $\left[0, 1\right]$.}
    \label{fig:original normalized load profile}
\end{figure}

The typical load patterns of the consumers are shown in Figure~\ref{fig:original normalized load profile}. It shows the normalized hourly mean heat load for each customer. This is the mean load for each respective hour and consumer for the whole year, subsequently normalized to the interval $\left[0, 1\right]$ such that the highest daily value for each consumer has value 1 and the lowest has value 0. Most consumers have peak loads in the morning, but individual peak hours vary between 5 and 10 am. Especially the smaller substations (the substations on the right side of the figure), which are mostly residential buildings, have an additional peak in the evening. The five larger substations (the left side of the figure) have low loads in the afternoon.

\subsection{Coordinated Optimal Load Shifting}
The first strategy for peak aggregate flow rate reduction is coordinated optimal load shifting. In this strategy, the consumers shift their load in time in a coordinated fashion, so as to reduce the maximum aggregate flow rate. We assume that consumers can only provide short-term flexibility and thus we constrain the total daily heat consumption of the consumers to be unaltered. Therefore, any load flexibility provided by a consumer at a given time is compensated either earlier or later the same day. For any given time $t$, the heat load $\dot{Q}_i(t)$ for consumer $i$ is given by
\begin{equation}
    \dot{Q}_i(t) = \rho \cdot c_p \cdot \Delta T_i(t) \cdot \dot{V}_i(t).
\end{equation}
where $\Delta T_i(t) = T_{\mathrm{SL},i}-T_{\mathrm{RL},i}$ is the difference between primary side supply and return temperature, $\Dot{V}_i$ is the volume flow rate, $\rho$ is the density of water assumed to be constant, and $c_p$ is the specific heat capacity of water. We make the assumption that the supply temperature is unaffected by us altering the flow rates of the system. In addition, we assume that $\Delta T$, and thus the return temperature, is also unaffected by altering the flow rate. Therefore, shifting the heat load corresponds to shifting the flow rate by the same factor.

For each hour, we bound the load the consumer can shift by a flexibility level which we denote $\alpha$. The altered individual hourly heat load may not deviate by a factor higher than $\alpha$ from the original heat load. Thus $\alpha$ becomes a metric on how much flexibility the consumers can provide. A small $\alpha$ may be possible to meet by leveraging the building's thermal inertia as a short-term storage. A large $\alpha$ on the other hand requires a large portion of the heat load to be flexible, which might require e.g. installation of local heat storage.

The load shifting strategy is applied to the dataset for each day solving the following optimization problem.
\begin{mini!} 
{\delta}{\max_{t = 1 \dots 24} \sum_{i=1}^n \delta_{i}(t) \cdot \dot{V}_i(t) }
{\label{eq:first opt problem}}{\label{eq:first opt cost}}
\addConstraint{1-\alpha \leq \delta_i(t) \leq 1+\alpha} \label{eq:first opt delta}
\addConstraint{\sum_{t=1}^{24} \delta_i(t) \cdot \dot{V}_i(t) \cdot \Delta T_i(t) = 0} \label{eq:first opt Q same}
\end{mini!}
where $\delta_i(t)$ corresponds to the portion of the original load which consumer $i$ shifts at hour $t$. $\delta_i(t) = 1$ corresponds to an unaltered flow rate, $\delta_i(t) = 0$ corresponds to completely stopping the flow rate, and $\delta_i(t) = 2$ corresponds to doubling it. The objective \eqref{eq:first opt cost} corresponds to the peak daily aggregate flow rate. Equation \eqref{eq:first opt delta} bounds the flexibility we allow each consumer by $\alpha$ and \eqref{eq:first opt Q same} restricts the daily heat load of each consumer to be unaltered. In theory, the solution to problem \eqref{eq:first opt problem} is non-unique. A simple example of this is that given that the aggregate flow rate is constant for a given day, then consumer 1 can increase their load at 6 am and reduce it at 7 am if consumer 2 performs an equal and opposite shift. This incurs the same cost as not shifting in the eyes of the optimizer, but clearly corresponds to different behaviors in the eyes of the consumers. Given different shifting-levels which provide equal peak flow rates, we prefer a low degree of load shifting. Hence we choose the solution to \eqref{eq:first opt problem} which additionally minimizes the following optimization problem.
\begin{mini!} 
{\delta}{\sum_{t=1}^{24}\sum_{i=1}^n |\delta_{i}(t)-1| }
{\label{eq:second opt cost}}{}
\addConstraint{1-\alpha \leq \delta_i(t) \leq 1+\alpha} \label{eq:second opt delta}
\addConstraint{\sum_{t=1}^{24} \delta_i(t) \cdot \dot{V}_i(t) \cdot \Delta T_i(t) = 0} \label{eq:second opt Q same}
\addConstraint{\max_{t = 1 \dots 24} \sum_{i=1}^n \delta_{i}(t)\cdot \dot{V}_i(t) \leq \dot{V}_*} \label{eq:second opt peak flow rate}
\end{mini!}
where $\dot{V}_*$ is the lowest peak aggregate flow rate found by solving \eqref{eq:first opt problem}. This corresponds to choosing the $\delta$ which minimizes the peak aggregate flow rate with the least total amount of load shifting.

In the case that the strategy is only implemented using flexibility from a few of the consumers, we assume that these consumers have knowledge of the expected aggregate flow rate of the system. Thus this scenario simply corresponds to the additional constraint $\delta_i(t)=1$ for all consumers $i$ which are not included in the strategy and all times $t$.

\subsection{Individual Return Temperature Limitation}
District heating suppliers generally require consumers to maintain a return temperature below the limit set by their technical connection requirement. The substations considered in the dataset were built in recent decades, so the design conditions have changed over the years as the district heating system has reduced its system temperatures. This results in different return temperature limits for each substation. These are documented in \ref{sec:appendix customers} together with the flow-weight rated return temperatures and further information on the substations. The measured return temperatures are often above the limits, caused by operation outside the design conditions or other faults on the secondary side (i.e. control setpoints, bypasses, faulty components).

In this strategy, we investigate the impact of strictly enforcing the return temperature limit, such that whenever a substation exceeds its return temperature limit in the dataset, we replace this with their maximum allowed temperature in the new data set. This increases the $\Delta T$ for the concerned data points and consequently leads to a reduction in volume flow rates $\dot{V}$. Assuming that the same heat load $\dot{Q}$ is delivered, the new volume flow rates are calculated according to equation \eqref{eq:nominal_volume_flow_rate}:
\begin{equation}
\label{eq:nominal_volume_flow_rate}
    \dot{V}_{i,\mathrm{new}}(t) = \frac{\dot{Q}_i(t)}{\rho \cdot c_p \cdot (T_{i,\mathrm{SL}}(t)-\max(T_{i,\mathrm{RL}}(t),T_{i,\mathrm{RL},\max}))}
\end{equation}
In practice, the behaviour described could be achieved by installing return temperature limiting valves to reduce the primary flow rates. However, this may result in lower secondary supply temperatures and consequently a reduced heat transfer if the heat exchanger of the substation is too small or the secondary equipment of the consumer is not operated optimally. Therefore, upgrades on the secondary side discussed previously in the literature review may be required before implementing the strategy to deliver the same amount of heat to the consumers.

\subsection{Individual Flow Rate Limitation}
Perhaps the most straight-forward approach to reducing the system flow rates is to dedicate a maximum flow rate $\dot{V}_{\text{limit},i}$ for each consumer $i$ which they are not allowed to exceed. For the remainder of this subsection, we will drop the subscript $i$ denoting a particular consumer $i$, as all of the following logic applies strictly to one consumer at a time. We set this limit to $\dot{V}_\text{limit} = (1-\beta) \max_t \dot{V}(t)$, where $\beta$ is the flow rate limitation level describing by which degree the peak flow rate is reduced, and $\max_t \dot{V}(t)$ is their individual peak flow rate for the full year of data. We also assume that whenever this limit renders a consumer unable to meet their original heat load, they will attempt to compensate for this at a later point, up to a horizon of 24 hours. Algorithm \ref{alg:described_algorithm} describes the process in detail. Here we first define a vector $Q_\text{deficit}$ as the heat load the consumer has not received in the last 24 hours due to the limitation on their flow rate. Then we iterate through all the hours of the year. The vector $Q_\text{deficit}$ is shifted, so that any heat load deficit older than 24 hours is discarded. Then we compare the flow rate $\Dot{V}(t)$ to the limit $\Dot{V}_\text{limit}$. If the original flow rate at hour $t$ was above $\Dot{V}_\text{limit}$, the new flow rate is reduced to $\Dot{V}_\text{limit}$ and the heat load which was not met is added to $Q_\text{deficit}$. If the flow rate at time $t$ does not exceed the limit, the consumer will increase their flow rate at that time by $\dot{V}_\text{compensated}$ corresponding to any heat load deficit in $Q_\text{deficit}$. If the deficit in $Q_\text{deficit}$ is greater than what can be delivered within the limit $\Dot{V}_\text{limit}$, the flow rate is set to $\Dot{V}_\text{limit}$. The additional heat $Q_\text{compensated}$ which is supplied this way is removed from the deficit vector $Q_\text{deficit}$. To avoid numerical issues, we avoid increasing the flow rate at hours where $\Delta T$ is below a threshold of 1 $^\circ$C, as division by $\Delta T$ is needed for this step, and numerical issues can occur when $\Delta T$ is close to 0.
\begin{algorithm}[H]
\caption{Flow Rate Limitation}
\label{alg:described_algorithm}
\begin{algorithmic}[1]
\State $Q_\text{deficit} \gets \left[0, 0, \dots, 0 \right]$ \Comment{heat deficit}
\For{$t = 1$ to $24 \cdot 365$}
    \For{ $i = 0$ to $ 23$} 
        \State $Q_\text{deficit}\left[i+1\right] \gets Q_\text{deficit}\left[i\right]$ 
    \EndFor
    \If{$\dot{V}(t) > \dot{V}_\text{limit}$}
        \State $Q_\text{deficit}\left[0\right] \gets (\dot{V}(t) - \dot{V}_\text{limit}) \cdot c_p \cdot \rho \cdot \Delta T(t)$
        \State $\dot{V}_{\text{new}}(t) \gets \dot{V}_\text{limit}$
    \Else
        \State $\dot{V}_{\text{compensated}} \gets \min \left( \frac{ \sum Q_\text{deficit}}{\rho \cdot c_p \cdot \Delta T(t)}, \dot{V}_\text{limit} - \dot{V}(t) \right)$
        \State $\dot{V}_{\text{new}}(t) \gets \dot{V}(t) + \dot{V}_{\text{compensated}}$
        \State $Q_\text{compensated} \gets \dot{V}_{\text{compensated}} \cdot \rho \cdot c_p \cdot \Delta T(t)$
        \For{ $h = 24$ to $1$}
            \State $Q_\text{hour} \gets \max{\left(Q_\text{compensated}, Q_\text{deficit}\left[h\right]\right)}$
            \State $Q_\text{deficit}\left[h\right] \gets Q_\text{deficit}\left[h\right] - Q_\text{hour}$
            \State $Q_\text{compensated} \gets Q_\text{compensated} - Q_\text{hour}$
        \EndFor
    \EndIf
\EndFor
\end{algorithmic}
\end{algorithm}

\subsection{Pumping Power and Aggregate Return Temperature}
Apart from the peak aggregate volume flow rate, we will investigate two additional performance indices; namely the \textit{required pumping energy} and the \textit{aggregate grid return temperature}.

We assume that the pumping power required to drive the flow rate through the system is proportional to the differential pressure at the pump, multiplied by the total flow rate \cite{thebible}. We then consider a model where the pressure difference $\Delta p$ at pump level is proportional to $\dot{V}^\lambda$ where $\dot{V}$ is the flow rate through the pump and $\lambda$ is an exponent. We investigate two cases for the exponent $\lambda$. Based on measured data at the pumping station in question from 2022, we consider $\lambda = 1.84$. We also consider $\lambda = 2$, which is the standard model to apply for flow rates through a pipe using the Darcy-Weisbach equation assuming turbulent flow. The ratio between the total yearly pumping power required to drive the system using any given strategy compared to the original data set is then given by
\begin{equation}
\label{eq:pumping power}
    \frac{\sum_{t=1}^{365 \cdot 24} \dot{V}_\text{strategy}(t)^{(1+\lambda)}}{\sum_{t=1}^{365 \cdot 24} \dot{V}_\text{original}(t)^{(1+\lambda)}}
\end{equation}
where $\dot{V}_\text{strategy}$ and $\dot{V}_\text{original}$ correspond to the aggregate volume flow rates using any of the strategies or the original data set.

To evaluate the impact on the aggregate grid return temperature, we evaluate the flow-rate-weighted aggregate return temperature $T_\mathrm{RL,aggregate,weighted}$ according to
\begin{equation}
    T_\mathrm{RL,aggregate,weighted} = \frac{\sum_{t=1}^{365 \cdot 24}\sum_{i=1}^{18} \dot{V}_{i}(t) \cdot T_{\mathrm{RL},i}(t) }{\sum_{t=1}^{365 \cdot 24} \dot{V}_\mathrm{aggregate}(t)}.
    \label{eq:weighted return temperature}
\end{equation}
This value corresponds to the mean aggregate return temperature in the system over the year where additional weight is granted to hours where the aggregate volume flow rate is high.

\section{Results}
\label{sec:results}
\subsection{Duration Curves}

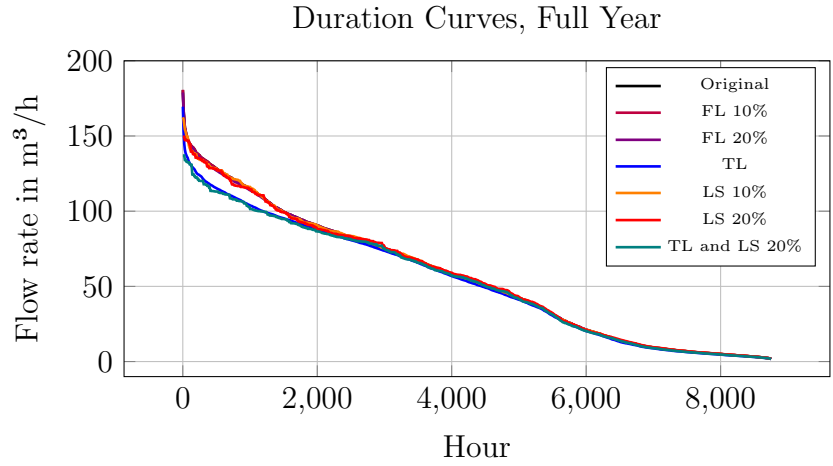
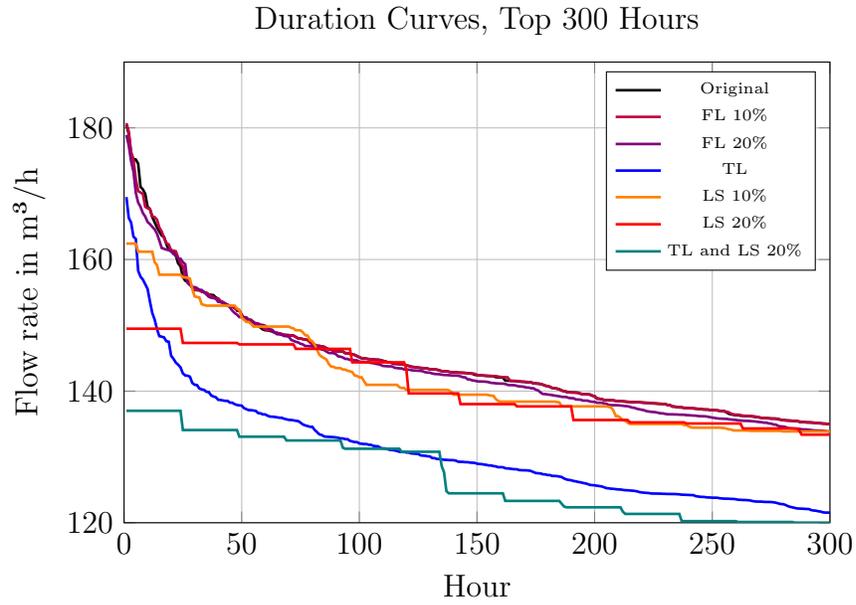
\begin{figure}[H]
    \centering
    
    \begin{subfigure}{0.8\textwidth} 
        \begin{tikzpicture}
            \begin{axis}[
                legend style={font=\tiny},
                width=\textwidth,
                height=0.3\textheight,
                ymin = -10, ymax = 200,
                grid=major,
                ylabel={Flow rate in m³/h},
                xlabel={Hour},
                title={Duration Curves, Full Year},
            ]
            \addplot[no markers, color=black, line width = 1.0pt] table[x=hour,y=original,col sep=comma]{data/duration_curves_every_10.csv};
                \addlegendentry{Original}
            
            \addplot[no markers, color=purple, line width = 1.0pt] table[x=hour,y=Flow rate limit 10,col sep=comma]{data/duration_curves_every_10.csv};
                \addlegendentry{FL 10\%}

            \addplot[no markers, color=violet, line width = 1.0pt] table[x=hour,y=Flow rate limit 20,col sep=comma]{data/duration_curves_every_10.csv};
                \addlegendentry{FL 20\%}
            
            \addplot[no markers, color=blue, line width = 1.0pt] table[x=hour,y=Limited return temperature,col sep=comma]{data/duration_curves_every_10.csv};
                \addlegendentry{TL}
            
            \addplot[no markers, color=orange, line width = 1.0pt] table[x=hour,y=Load shift 10,col sep=comma]{data/duration_curves_every_10.csv};
                \addlegendentry{LS 10\%}

            \addplot[no markers, color=red, line width = 1.0pt] table[x=hour,y=Load shift 20,col sep=comma]{data/duration_curves_every_10.csv};
                \addlegendentry{LS 20\%}
            
            \addplot[no markers, color=teal, line width = 1.0pt] table[x=hour,y=Lim. return temp. and load shift 20,col sep=comma]{data/duration_curves_every_10.csv};
                \addlegendentry{TL and LS 20\%}
                   
            \end{axis}
        \end{tikzpicture}
        \caption{The full year of 2022.}
        \label{fig:duration curves full year}
    \end{subfigure}
    
    \begin{subfigure}{0.8\textwidth} 
        \begin{tikzpicture}
            \begin{axis}[
                legend style={font=\tiny},
                width=\textwidth,
                height=0.4\textheight,
                xmin=0, xmax=300,
                ymin = 120, ymax = 190,
                grid=major,
                ylabel={Flow rate in m³/h},
                xlabel={Hour},
                title={Duration Curves, Top 300 Hours},
            ]
            \addplot[no markers, color=black, line width = 1.0pt] table[x index=0,y=original,col sep=comma]{data/duration_curves.csv};
                \addlegendentry{Original}
            
            \addplot[no markers, color=purple, line width = 1.0pt] table[x index=0,y=Flow rate limit 10,col sep=comma]{data/duration_curves.csv};
                \addlegendentry{FL 10\%}

            \addplot[no markers, color=violet, line width = 1.0pt] table[x index=0,y=Flow rate limit 20,col sep=comma]{data/duration_curves.csv};
                \addlegendentry{FL 20\%}
            
            \addplot[no markers, color=blue, line width = 1.0pt] table[x index=0,y=Limited return temperature,col sep=comma]{data/duration_curves.csv};
                \addlegendentry{TL}
            
            \addplot[no markers, color=orange, line width = 1.0pt] table[x index=0,y=Load shift 10,col sep=comma]{data/duration_curves.csv};
                \addlegendentry{LS 10\%}

            \addplot[no markers, color=red, line width = 1.0pt] table[x index=0,y=Load shift 20,col sep=comma]{data/duration_curves.csv};
                \addlegendentry{LS 20\%}
            
            \addplot[no markers, color=teal, line width = 1.0pt] table[x index=0,y=Lim. return temp. and load shift 20,col sep=comma]{data/duration_curves.csv};
                \addlegendentry{TL and LS 20\%}
                
            \end{axis}
        \end{tikzpicture}
        \caption{The top 300 hours of 2022.}
        \label{fig:duration curves zoomed in}
    \end{subfigure}
    
    \caption{Duration curves of total flow rates using flow rate limitation (FL), return temperature limitation (TL) or coordinated load shifting (LS). The results are displayed over the full year (\ref{fig:duration curves full year}) and detailed over only the top 300 hours (\ref{fig:duration curves zoomed in}).}
    \label{fig:duration curves}
\end{figure}

Figure \ref{fig:duration curves} displays the duration curves, i.e. the values sorted in descending order, for aggregate volume flow rates in the system with the different considered strategies. These duration curves indicate what the flow rates are at the peak hours of the year. One primary note is that the individual flow rate limitation (purple and violet lines) hardly affects the curve in comparison to the original data, even when reducing the peak flow rate by 20 \%. However, both the return temperature limitation (blue line) and the load coordination strategies (orange and red lines) have a significant impact on the peak aggregate flow rate. The load coordination strategies in particular provide the most significant impact during the 30 peak delivery hours of the year. After that point, the difference between the load coordination strategy and the original data becomes smaller. The limited return temperature strategy on the other hand acts differently. The effect at the 30 peak hours of the year is less noticeable compared to the load coordination strategy. However, the flow rates are reduced over a wider range of hours. The reduction in flow rates is noticeable beyond the first 1000 peak hours of the year. This is not an unexpected result, given that the return temperature limitation strategy corresponds to a strict upgrade of the consumer substations, whereas the load coordination strategy merely shifts the load away from the peak hours. We also demonstrate the result of first applying a return temperature limitation, and then performing load coordination on top of this (teal line). It is interesting to see that the strategies appear to complement each other, such that the benefits of reduced flow rates granted by combining the strategies is similar to the sum of the benefits of each individual strategy. In the original data set the peak flow rate was 180 m$^3$/h, under 20 \% load shift it was 149 m$^3$/h, under return temperature limitation it was 169 m$^3$/h and under the combined strategy it was 137 m$^3$/h. Hence our study indicates that one does not have to consider them exclusive from one another for improving system performance.

\subsection{Partial Strategy Implementation}
\label{sec:partial_strategy_implementation}
In this section we consider the effect of implementing the strategies with only a subset of the consumers. The consumers chosen for inclusion are selected greedily, such that we iteratively include in the strategy the consumer that provides the greatest reduction in aggregate peak flow rate. Thus the order of the included substations is different from Figure~\ref{fig:original_load_bars} and may also differ between the three strategies.

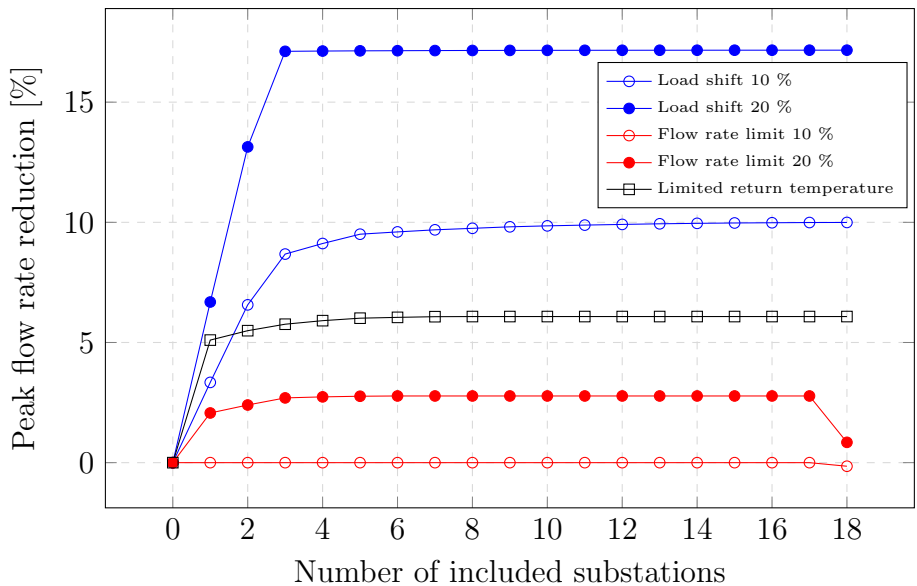
\begin{figure}[H]
    \centering
    \begin{tikzpicture}
    \begin{axis}[
        width=0.9\linewidth, 
        height=0.6\linewidth,
        grid=major, 
        grid style={dashed,gray!30},
        xlabel={Number of included substations}, 
        ylabel={Peak flow rate reduction [\%]},
        legend style={at={(0.8,0.6)}, anchor=south, font=\tiny},
        legend cell align={left}
    ]
        \addplot[color=blue, mark=o] table[x=meters,y=Load shift 10, col sep=comma] {data/included_meters.csv};
        \addplot[color=blue, mark=*] table[x=meters,y=Load shift 20, col sep=comma] {data/included_meters.csv};
        \addplot[color=red, mark=o] table[x=meters,y=Flow rate limit 10, col sep=comma] {data/included_meters.csv};
        \addplot[color=red, mark=*] table[x=meters,y=Flow rate limit 20, col sep=comma] {data/included_meters.csv};
        \addplot[color=black, mark=square] table[x=meters,y=Limited return temperature, col sep=comma] {data/included_meters.csv};

        \legend{
            Load shift 10 \%,
            Load shift 20 \%,
            Flow rate limit 10 \%,
            Flow rate limit 20 \%,
            Limited return temperature
        }
    \end{axis}
\end{tikzpicture}
    \caption{The effect on peak yearly flow-rate when including a larger number of substations.}
    \label{fig:included meters}
\end{figure}

Figure~\ref{fig:included meters} shows the effect of this. Consider first the flow rate limitation strategy. At 10~\% reduction, we find that there is little benefit to imposing a limit. At 20~\% we see a rather interesting effect. Adding the first consumer provides around 2~\% reduction in aggregate peak flow rate. However, the addition of the final consumer actually reduces this benefit. This is explained by our assumption that consumers are expected to compensate for lost heat load. What can happen is that by limiting the heat load of a consumer at the time of their peak consumption, they can shift this load to a later time when it overlaps poorly with the aggregate flow rates, generating a higher overall aggregate peak. Such is the case when we add the last consumer to the strategy. An important note here is that we have the greedy policy for including consumers into this strategy actually makes the highlighted results the best case scenario. 

When considering load coordination, we find a rather predictable result when allowing for 10~\% flexibility level. In the first three additions of actors, we reach a significant improvement. In consideration of our data set, this makes sense, as the load is dominated by a few large consumers. Hence adding these consumers first will provide the most improvement. We then see the improvement converging to 10~\% as more and more consumers are added. We do not expect to be able to exceed 10~\% improvement, as at most 10~\% of the flow rate can be shifted at any time. When we consider the greater flexibility level of 20~\%, we see that we do not converge to 20~\% improvement. Rather, the improvement plateaus at around 18~\% after including three consumers into the strategy. Hence, in our data set which is dominated by large consumers, it makes sense to first consider including only the largest consumers. One should then ensure that they are able to provide significant flexibility through for instance installing local short-term storage.

Finally, we consider the effects of limiting return temperatures, and see that the majority of the improvement comes from one consumer with a large consumption and poor equipment. Similarly to the load coordination strategy case, a targeted intervention for one key consumer provides the most significant improvement.

\subsection{Impact of Flexibility or Limitation Level}
In this section we investigate the impact the the flexibility level $\alpha$ and the flow rate limitation level $\beta$ have on the aggregate peak flow rate under the coordinated load shifting strategy and the flow rate limitation strategies respectively.
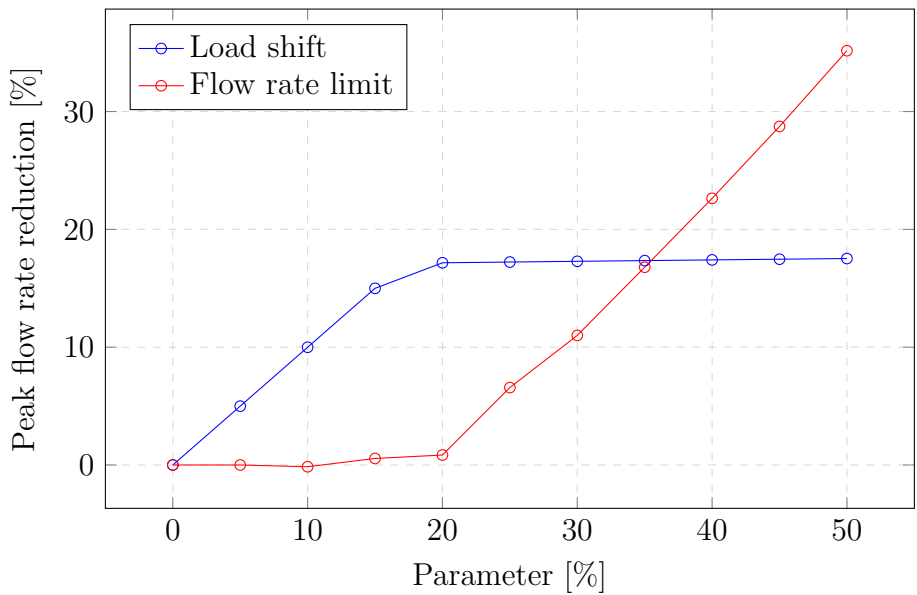
\begin{figure}[H]
    \centering
    \begin{tikzpicture}
    \begin{axis}[
        width=0.9\linewidth, 
        height=0.6\linewidth,
        grid=major, 
        grid style={dashed,gray!30},
        xlabel={Parameter [\%]}, 
        ylabel={Peak flow rate reduction [\%]},
        legend pos=north west,
        legend cell align={left}
    ]
        \addplot[color=blue, mark=o] table[x=percentages,y=optimal load coordination, col sep=comma] {data/percentage_parameter.csv};
        \addplot[color=red, mark=o] table[x=percentages,y=flow rate limitation, col sep=comma] {data/percentage_parameter.csv};
        \legend{
            Load shift,
            Flow rate limit,
        }
    \end{axis}
\end{tikzpicture}
    \caption{Reduction in the yearly peak flow rate, when imposing a larger flexibility level $\alpha$ (Load shift) or peak flow rate limit $\beta$ (Flow rate limit).}
    \label{fig:flexibility level and flow rate limit level}
\end{figure}

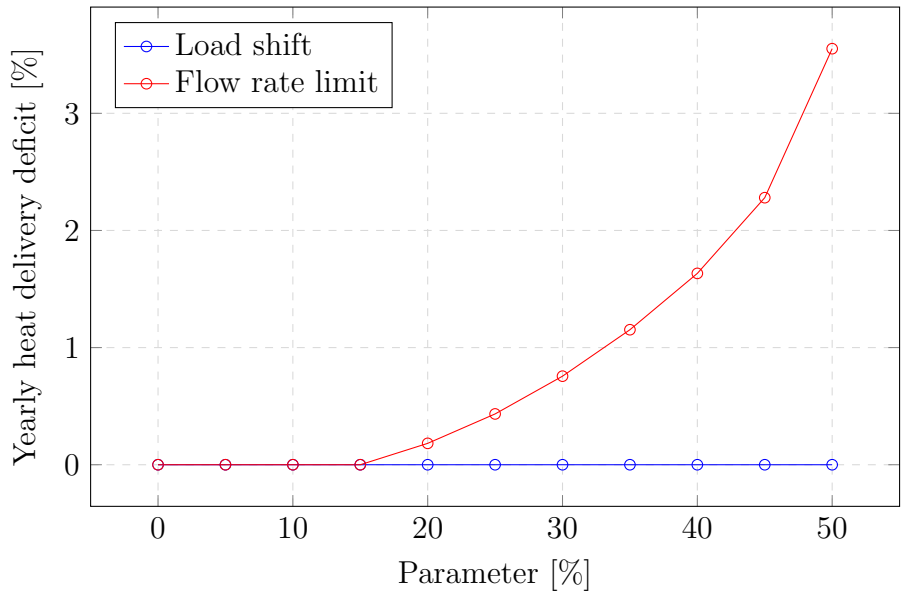
\begin{figure}[H]
    \centering
    \begin{tikzpicture}
    \begin{axis}[
        width=0.9\linewidth, 
        height=0.6\linewidth,
        grid=major, 
        grid style={dashed,gray!30},
        xlabel={Parameter [\%]}, 
        ylabel={Yearly heat delivery deficit [\%]},
        legend pos=north west,
        legend cell align={left}
    ]
        \addplot[color=blue, mark=o] table[x=percentages,y=optimal load coordination deficit, col sep=comma] {data/percentage_parameter.csv};
        \addplot[color=red, mark=o] table[x=percentages,y=flow rate limitation deficit, col sep=comma] {data/percentage_parameter.csv};
        \legend{
            Load shift,
            Flow rate limit,
        }
    \end{axis}
\end{tikzpicture}
    \caption{Reduction in the total heat supplied to all the buildings, when imposing a larger flexibility level $\alpha$ (Coordinated load shift) or flow rate limitation level $\beta$ (Flow rate limit).}
    \label{fig:heat load deficit}
\end{figure}

 Figure \ref{fig:flexibility level and flow rate limit level} shows the effect of the choice of these parameters on the peak yearly aggregate flow rate. As we noticed in the previous section, the load coordination strategy stagnates at 18 \% improvement after introducing around the same level of flexibility $\alpha$. The flow rate limit on the other hand shows an interesting result. When the limit-level $\beta$ is below 20 \%, we see no significant improvement. However, after 20 \% there appears to be improvement which is linear in $\beta$. An explanation for this can be found in Figure \ref{fig:heat load deficit}, which shows the reduction of total yearly heat load. As expected, there is no reduction of heat load in the coordinated load shifting case as this is a constraint in the optimization \eqref{eq:first opt problem}. In the flow rate limit case however, we see that after a 15 \% limit $\beta$, we are no longer able to meet the desired heat load. At $\beta = 35$ \%, which is when the flow rate limitation strategy provides the same improvement as the load coordination strategy, we find that the total yearly heat load is reduced by slightly more than 1 \%. Thus at this point we may expect that the strategy starts to impact consumer comfort due to a deficit in heat load.

\subsection{Additional Benefits}
Our results so far concern the impact our different strategies impose on the peak yearly aggregate flow rate in the system. However, the introduction of one or more of these strategies can also provide other effects and benefits which impact the choice of which method to implement. While a detailed investigation is outside the scope of this paper, we will briefly see what effect the different methods will have on two important factors: The estimated required pumping energy and the aggregate return temperature.

In Table \ref{tab:pumping power} we show the percentage of original yearly pumping energy required to drive the flow rate under varying strategies, calculated through equation \eqref{eq:pumping power}. The choice of the exponent $\lambda$ appears to make only a little difference in the ratios. We can note that while both the load coordination strategy and the flow rate limitation strategy provide non-significant improvements, the limited return temperature strategy provides a major improvement of almost 20 \% reduction in required pumping power. We attribute this to the earlier observation regarding duration curves. Namely that the limited return temperature strategy reduces the system flow rates over more than 1000 hours. The flow rate limitation strategy on the other hand hardly impacts the aggregate flow rates and the load coordination strategy only impacts the first 30 peak hours of the year.

\begin{table}
\centering
\caption{Pumping power ratios for the different strategies coordinated load shifting (LS),  flow rate limitation (FL) and return temperature limitation (TL).}
\begin{tabular}{l|ccccc}
 $\lambda$ &  LS 10\% &  LS 20\% &  FL 10\% &  FL 20\% &  TL \\ \hline
    1.84 &          0.990 &          0.969 &          1.0 &        0.986 &      0.818 \\
    2.00 &          0.988 &          0.966 &          1.0 &        0.985 &      0.806 \\
\end{tabular}
\label{tab:pumping power}
\end{table}

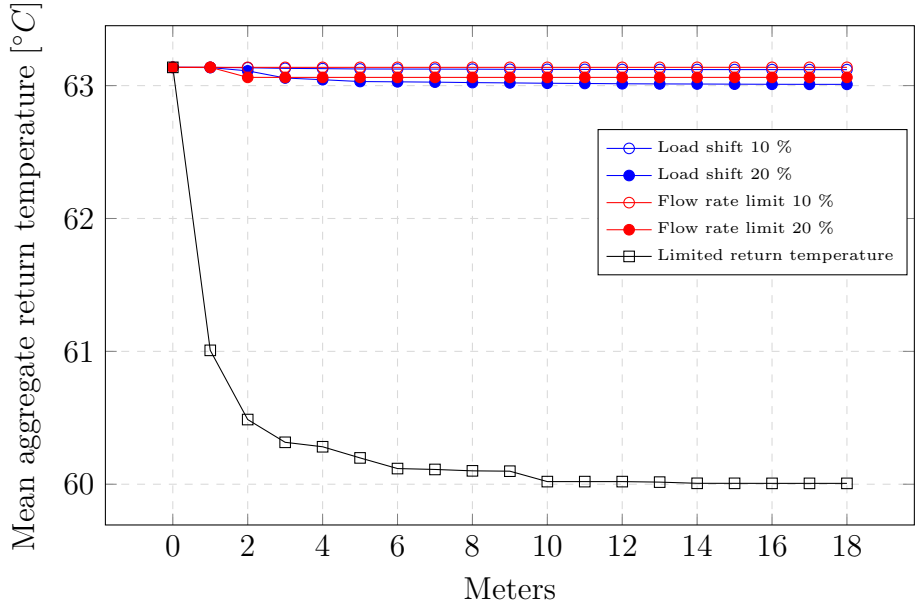
\begin{figure}[H]
    \centering
    \begin{tikzpicture}
    \begin{axis}[
        width=0.9\linewidth, 
        height=0.6\linewidth,
        grid=major, 
        grid style={dashed,gray!30},
        xlabel={Meters}, 
        ylabel={Mean aggregate return temperature [$^\circ C$]},
        legend style={at={(0.8,0.5)}, anchor=south, font=\tiny},
        legend cell align={left}
    ]
        \addplot[color=blue, mark=o] table[x=meters,y=Load shift 10, col sep=comma] {data/return_temperatures.csv};
        \addplot[color=blue, mark=*] table[x=meters,y=Load shift 20, col sep=comma] {data/return_temperatures.csv};
        \addplot[color=red, mark=o] table[x=meters,y=Flow rate limit 10, col sep=comma] {data/return_temperatures.csv};
        \addplot[color=red, mark=*] table[x=meters,y=Flow rate limit 20, col sep=comma] {data/return_temperatures.csv};
        \addplot[color=black, mark=square] table[x=meters,y=Limited return temperature, col sep=comma] {data/return_temperatures.csv};

        \legend{
            Load shift 10 \%,
            Load shift 20 \%,
            Flow rate limit 10 \%,
            Flow rate limit 20 \%,
            Limited return temperature,
        }
    \end{axis}
\end{tikzpicture}
    \caption{The impact on the annual weighted mean return temperature in the grid by imposing strategies upon an increasing number of consumers.}
    \label{fig:impact_TRL_limit}
\end{figure}

The second benefit we investigate is reduced aggregate return temperature in the grid branch. Analogous to the partial strategy implementation discussed in section~\ref{sec:partial_strategy_implementation}, Figure~\ref{fig:impact_TRL_limit} shows the impact of the strategies on the annual weighted aggregate return temperature given by equation \eqref{eq:weighted return temperature} by adding more substations to each strategy. If all of the substations are included in the return temperature limitation strategy, the aggregate return temperature in the grid branch is reduced by 3~K from 63~\degree C to 60~\degree C. However, this is already achieved by including eight substations in the strategy and only the first substation accounts for 2~K of the reduction. This confirms the finding from section~\ref{sec:partial_strategy_implementation} that targeting only a few key consumers has the greatest impact. None of the other considered strategies have a significant impact on the return temperature.

\subsection{Discussion}
The results related to coordinated load shifting and flow rate limitation rely on the assumption that $\Delta T$ is unaltered by a change in flow rate. In practice, this is unlikely to be the case. However, while there should be a slight correlation between volume flow rates and $\Delta T$, it is important to note that we consider only small changes in flow rates in the magnitude of 10 to 20 \%. Hence the effect on $\Delta T$ should also be minor. Additionally, to fully incorporate the effect of this correlation in the study, we would need either much more sophisticated models or data measurements from the secondary side of the substations, which we do not have available.

An error that this assumption of unaltered $\Delta T$ could potentially induce is the following. When performing load shifting, we assume that $\dot{Q} \propto \Delta T \cdot \dot{V}$. Hence, to deliver a constant daily heat load while minimizing flow rates, it would make sense to shift the heat load to a time of day when $\Delta T$ is large, thereby minimizing the required $\dot{V}$. Such a strategy relies critically on the previous assumption that $\Delta T$ is unaltered by this shift. However, this would imply a reduction in the total yearly volume flow. When we compare the total yearly volume flow when considering a flexibility level of 10 or 20 \%, we see a reduction in total yearly volume flow of 0.04 or 0.24 \% respectively, leading us to believe that this is not the main factor behind the peak flow rate reduction that we see.

The load shifting strategy was shown to be the method with the highest impact on the yearly peak load. However, this method critically requires loads which are shiftable in time. This holds for instance if a large part of the peak load is for space heating purposes. However, in our data set, we see that adding the largest consumers to the strategy first has the most impact and in this case these consumers are industrial consumers. Hence the ability to shift their heat loads depends fully on what sort of operations they are running. Hence in order to shift their load as proposed in our work, one may need to install short term heat storage which constitutes a cost.

The return temperature limitation strategy assumes that some substations may need to be upgraded to ensure that the same heat load can be supplied. These upgrades could additionally reduce return temperatures at times when the set limit is not exceeded, resulting in a further reduction in the mean aggregate return temperature and the pumping power ratios. However, it is necessary to analyse and model all substations in detail to investigate this effect.

Finally, it should be noted that the results are tied to the dataset used in this work and the applicability to other district heating systems should be further investigated. However, the subset of district heating substations already includes three different types of consumers (residential, commercial, industrial) and their characteristics are assumed to be common in existing urban high-temperature district heating grids. The conclusion that targeting a few large key consumers with the proposed strategies is sufficient to achieve most of the benefits for the grid is likely to hold true for many district heating systems with similar characteristics and a mixed consumer structure.

\section{Conclusion}
\label{sec:conclusion}
In this paper we investigated the impact of imposing three different strategies for reducing peak aggregate flow rates on a data set. The concerned strategies were optimal coordinated load shifting, return temperature limitation and flow rate limitation. We found that imposing coordinated load shifting had the greatest impact on the peak system flow rate, followed by return temperature limitation. Additionally, we note that having only a few large consumers providing flexibility for the rest of the grid is approximately as efficient as coordinating among the whole consumer population. This can be explained in part by the fact that our data set had a few major consumers which dominated the system behavior. Additionally, we noted that there is a diminishing return on the level of flexibility imposed on the consumers, such that allowing for more than 18 \% load flexibility each hour incurs no further reductions in aggregate peak flow rates.

A fairly surprising result is that imposing flow rate limits on consumers can give rise to higher flow rate peaks. This result hinges on our assumption that consumers compensate for any heat load deficit incurred by the flow rate limitation at later times.

While load coordination outperformed return temperature limitation in terms of peak flow rate reduction, the return temperature limitation seemingly provides additional benefits, in the form of an overall lower requirement on pumping energy as well as lower aggregate return temperatures.

\subsection{Future Work}
This work concerns only the expected impact a certain strategy might have on our system behavior. The important question of how these strategies should be implemented in practice is part of future work, and should be expected to affect the results.

The results should also be confirmed on further datasets with different sizes and properties. For instance, one of the major findings of our work is that a coordinated load shifting strategy can be assigned to only the largest consumers of the system with diminishing rewards after that point. In a data set with a higher proportion of small residential customers, this conclusion may no longer hold. The dominance of the large consumers also implies that a detailed investigation of the larger substations in the dataset and their secondary side infrastructure is needed to enable lower temperatures in the district heating system. Additionally, we consider a rather old German district heating system, with supply and return temperatures which are disproportionately high to systems one expects to find in many other regions.

\section*{CRediT authorship contribution statement}
\textbf{Felix Agner:} Conceptualization, Methodology, Software, Writing - Original Draft, Visualization

\textbf{Ulrich Trabert:} Conceptualization, Methodology, Software, Data Curation, Writing - Review \& Editing, Visualization

\textbf{Anders Rantzer:} Writing - Review \& Editing, Supervision, Project Administration, Funding Aquisition

\textbf{Janybek Orozaliev:} Writing - Review \& Editing, Supervision, Project Administration, Funding Aquisition

\section*{Declaration of competing interest}
The authors declare that they have no known competing
financial interests or personal relationships that could have
appeared to influence the work reported in this paper.

\section*{Data availability}
The data used in this work is confidential and therefore cannot be published.

\section*{Acknowledgements}
\textbf{Felix Agner and Anders Rantzer:} This work is funded by the European Research Council (ERC) under the European Union's Horizon 2020 research and innovation program under grant agreement No 834142 (ScalableControl). The authors are members of the ELLIIT Strategic Research Area at Lund University.\\
\textbf{Ulrich Trabert and Janybek Orozaliev:} This work was supported by the German Federal Ministry for Economic Affairs and Climate Action [Funding Code: 03EN3023A].

\appendix

\section{Substation information}
\label{sec:appendix customers}
Table \ref{tab:customer return temperature information} details the mean and maximum heat loads and return temperature for each consumer. It also shows the return temperature limitation set by the utility and a label for each consumer dictating if they are residential, commercial or industrial.
\begin{table}[H]
    \centering
    \caption{Heat load and return temperature information from 2022 and meta-data for each substation, sorted by peak heat load.}
    \begin{tabular}{ccccccc}
        \makecell{Meter\\ID} & \makecell{$\dot{Q}_{\mathrm{max}}$\\in kW} & \makecell{$\dot{Q}_{\mathrm{mean}}$\\in kW} & \makecell{$T_{\mathrm{RL,mean}}$\\in \degree C} & \makecell{$T_{\mathrm{RL,max}}$\\in \degree C} & \makecell{$T_{\mathrm{RL,limit}}$\\in \degree C} & \makecell{Consumer\\type} \\ \hline
        1 & 5356 & 1276 & 58.4 & 66.9 & 60 & Industrial \\
        2 & 4799 & 953 & 63.9 & 74.6 & 75 & Industrial \\
        3 & 2012 & 890 & 74.9 & 94.0 & 65 & Commercial\\
        4 & 1164 & 210 & 61.0 & 67.0 & 50 & Commercial\\
        5 & 590 & 179 & 59.5 & 75.2 & 65 & Residential\\
        6 & 188 & 50 & 52.4 & 93.4 & 50 & Residential\\
        7 & 174 & 40 & 53.0 & 68.3 & 50 & Commercial\\
        8 & 162 & 51 & 42.8 & 64.3 & 55 & Residential\\
        9 & 111 & 30 & 60.2 & 78.6 & 50 & Commercial\\
        10 & 83 & 27 & 47.2 & 62.6 & 65 & Residential\\
        11 & 60 & 19 & 47.2 & 69.8 & 65 & Residential\\
        12 & 55 & 14 & 48.2 & 61.2 & 50 & Residential\\
        13 & 45 & 16 & 70.6 & 102.0 & 50 & Residential\\
        14 & 44 & 10 & 46.1 & 55.9 & 65 & Residential\\
        15 & 38 & 10 & 53.7 & 62.1 & 50 & Residential\\
        16 & 27 & 8 & 53.9 & 93.1 & 50 & Residential\\
        17 & 19 & 7 & 46.5 & 87.3 & 50 & Residential\\
        18 & 11 & 2 & 41.6 & 87.0 & 65 & Commercial\\
    \end{tabular}
    \label{tab:customer return temperature information}
\end{table}

\section*{Declaration of Generative AI and AI-assisted technologies in the writing process}
During the preparation of this work the authors used 'DeepL - Write' in order to check grammar and improve readability of the text. After using this service, the authors reviewed and edited the content as needed and take full responsibility for the content of the publication.


\bibliographystyle{model3-num-names}
\bibliography{bibliography}





\end{document}